\documentstyle[epsfig]{aipproc}
\def\prl{Phys. Rev. Lett.}
\def\prb{Phys. Rev. B}
\begin{document}
\title{1/8 Doping Anomalies and Oxygen Vacancies in Underdoped Superconducting Cuprates}
\author{Joshua L. Cohn}
\address{Department of Physics, University of Miami, Coral Gables, FL 33124}

\maketitle

\begin{abstract}
Measurements of thermal conductivity ($\kappa$) versus temperature and doping
for several cuprate superconductors are discussed.  Suppressed values of the
normal-state $\kappa$ and the slope change in $\kappa$ at $T_c$, observed near
1/8 doping in both YBa$_2$Cu$_3$O$_{6+x}$ and Hg cuprates, are attributed to
local lattice distortions and the suppression of the superconducting
condensate, respectively.  Both phenomena are proposed to arise from small
domains of localized planar holes, presumably a manifestation of phase
separation.  It is suggested that the phase behavior of $\kappa$ reflects
stripe dynamics and the 1/8 doping anomalies stripe pinning by oxygen-vacancy
clusters.
\end{abstract}

\section*{Introduction}
There is mounting experimental evidence that novel charge- and spin-ordered
phases are generic to underdoped cuprates\cite{Cuprates}. The doped holes in
Nd-doped La$_{1-x}$Sr$_x$CuO$_4$ segregate into periodic stripes that separate
antiferromagnetically ordered, hole-poor domains.  Lattice distortions pin
these stripes, and this pinning is most effective for planar hole
concentrations near $p$=1/8 where the stripe modulation wavelength is
commensurate with the lattice.  In the absence of pinning, stripe modulations
are presumed to be fluctuating and/or disordered, and this is the emerging
picture for La$_{1-x}$Sr$_x$CuO$_4$ (La-214) and YBa$_2$Cu$_3$O$_{6+x}$
(Y-123)\cite{DynamicStripes}.  We have recently demonstrated, through
measurements of thermal conductivity ($\kappa$)\cite{CohnHg}, that both Y-123
and HgBa$_2$Ca$_{m-1}$Cu$_m$O$_{2m+2+\delta}$ [Hg-12($m$-1)$m$, $m$=1, 2, 3]
exhibit doping anomalies near $p$=1/8 that can be attributed to the presence
of localized charge and associated lattice distortions.  Here we discuss the
oxygen doping behavior of the anomalies in Hg cuprates and present new results
on vacancy-free Ca-doped YBa$_2$Cu$_4$O$_8$ (Y-124) that suggest elastic
properties may reflect stripe dynamics and stripe fragments may pin near
oxygen-vacancy clusters.
\begin{figure}[t!] 
\centerline{\epsfig{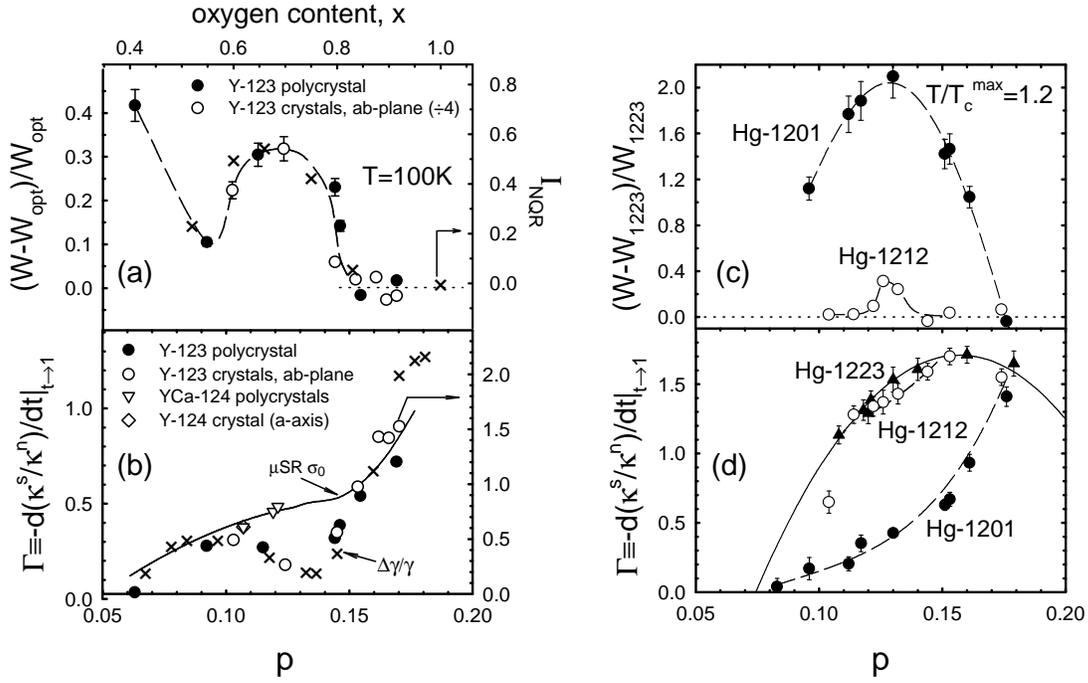}}%
\vskip -3.4in
\caption{(a) Thermal resistivity at $T$=100K
relative to that at $p_{opt}=0.16$ for Y-123 polycrystals and the
ab-plane of single crystals \protect\cite{Popoviciu} [$W_{opt}=0.21$ mK/W
(0.08mK/W) for the polycrystal (crystals)]. The hole concentration
per planar Cu atom, $p$, was determined from thermopower
measurements and the empirical relations with bond valence sums
59 established by Tallon {\it et al.}\protect\cite{PfromTEP}.  Also plotted
($\times$'s) is the relative intensity of anomalous $^{63}$Cu NQR
signals\protect\cite{HamScal}. The dashed curve is a guide to the eye. (b)
The normalized slope change in $\kappa(T)$ at $T_c$ {\it vs}
doping for each of the Y-123 specimens in (a), for the a-axis of
single-crystal Y-124\protect\cite{Cohn124xtals}, and for polycrystal
Y$_{1-x}$Ca$_x$-124\protect\cite{CohnYCa124}. The Y$_{1-x}$Ca$_x$-124 data
are multiplied by 2.3 so that they coincide with the Y-124 crystal
data at x=0. Solid (open) symbols are referred to the left (right)
ordinate. Also shown ($\times$'s) are the normalized electronic
specific heat jump\protect\cite{Loram}, $\Delta\gamma/\gamma$, and (solid
curve) the $\mu$SR depolarization rate\protect\cite{muSRY123} (in
$\mu$s$^{-1}$), divided by 1.4 and 2.7, respectively, and
referred to the right ordinate. (c) Thermal resistivity for
Hg-1201 and Hg-1212 polycrystals relative to that of
Hg-1223\protect\cite{CohnHg}. Dashed curves are guides. (d) The normalized
slope change in $\kappa(T)$ at $T_c$ {\it vs} doping for Hg
cuprates.  The solid line is $1.71-250(p-0.157)^2$.  Dashed curves
are guides.}
\label{fig1}
\end{figure}

The $p$=1/8 features (Fig.~\ref{fig1}) are evident in the doping behavior of
the normal-state thermal resistivity, $W=1/\kappa$, and the normalized change
in temperature derivative of $\kappa$ that occurs at $T_c$, $\Gamma\equiv
-d(\kappa^s/\kappa^n)/dt|_{t=1}$ [$t=T/T_c$ and $\kappa^s (\kappa^n)$ is the
thermal conductivity in the superconducting (normal) state].  For Y-123, there
is a compelling correlation of $W(p)$ and $\Gamma(p)$ with the doping behavior
of anomalous $^{63}$Cu NQR spectral weight\cite{HamScal}, attributed to
localized holes, and the electronic specific heat jump\cite{Loram},
$\Delta\gamma/\gamma$, respectively [crosses in Fig.'s 1 (a) and (b)]. Thus
$W$ probes lattice distortions associated with localized holes, and $\Gamma$
the change in low-energy spectral weight induced by superconductivity.

Since both $\Gamma$ and $\Delta\gamma/\gamma$ provide bulk measures of the
superfluid volume, it is significant that the muon spin rotation ($\mu$SR)
depolarization rate [$\sigma_0$ in Fig. 1 (b)], proportional to the superfluid
density, exhibits no anomalous behavior near 1/8 doping.  The $\mu$SR signal
originates in regions of the specimen where there is a flux lattice. The
apparent discrepancy between $\sigma_0$ and $\Gamma$ or $\Delta\gamma/\gamma$
is resolved if the material is inhomogeneous, composed of non-superconducting
clusters embedded in a superconducting network.  The suppression of $\Gamma$
and $\Delta\gamma/\gamma$ below the scaled $\sigma_0$ curve in Fig. 1 (a) are
then measures of the non-superconducting volume fraction. Taken together, the
$W$ and $\Gamma$ data imply that the non-superconducting regions are comprised
of localized holes and associated lattice distortions, i.e. polarons.  It is
plausible that these hole-localized regions are stripe domains akin to those
inferred from neutron scattering\cite{Cuprates}.  That they produce lattice
thermal resistance implies they are static on the timescale of the average
phonon lifetime, estimated as a few ps in the ab-plane of Y-123\cite{CohnHg}.
Given that $T_c$ is not substantially suppressed near $p$=1/8 implies that
these regions do not percolate, and thus the picture is one of small clusters
of localized holes and their associated lattice distortions, perhaps no larger
than the stripe unit cell ($2a\times8a$ where $a$ is the lattice
constant\cite{Cuprates}), separated by a distance comparable to the phonon
mean free path (about 100${\rm\AA}\sim 25a$ at 100K).


For the Hg materials the $p$=1/8 enhancement of $W$ and suppression of
$\Gamma$ is most prominent in single-layer Hg-1201, less so in double-layer
Hg-1212, and absent or negligible in three-layer Hg-1223.  This trend follows
that of the oxygen vacancy concentration: a single HgO$_{\delta}$ layer per
unit cell contributes charge to $m$ planes in Hg-12($m$-1)$m$ so that the
oxygen vacancy concentration, $1-\delta$, increases with decreasing $m$
\cite{Occupancy}.  The absence of suppression in $\Gamma$ near 1/8 doping for
Hg-1223 [Fig.~1~(d)] suggests that this material has sufficiently few
localized-hole domains that their effects in $W$ and $\Gamma$ are
unobservable.  Thus we employ the Hg-1223 $W(p)$ data as a reference and plot
the differences for the other two compounds in Fig. 1 (c). Comparing
Fig.'s~1~(c) and (d) we see that for both Hg-1201 and Hg-1212 $W$ is enhanced
and $\Gamma$ is suppressed relative to values for Hg-1223 in common ranges of
$p$, with maximal differences near $p$=1/8.

\begin{figure}[t!] 
\vskip -.5in \centerline{\epsfig{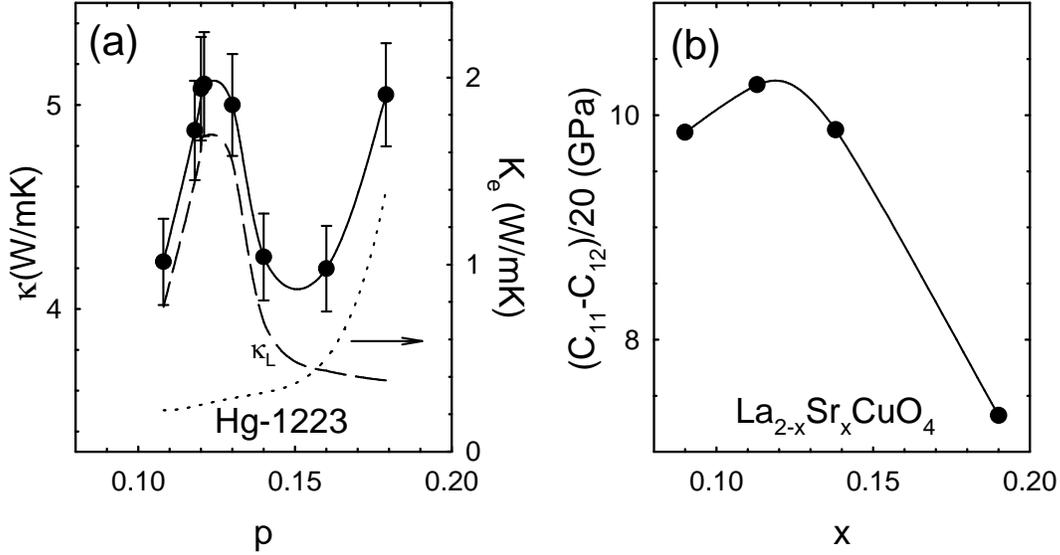}}
\vskip -3in
\caption{(a) Thermal conductivity of Hg-1223 polycrystal versus doping.
The dotted and dashed curves are proposed electronic (right ordinate) and
lattice thermal conductivities, respectively.  The solid line is a guide to
the eye. (b) sound velocity data for single-crystal La-214 from
Ref.~\protect\cite{LSCOelastic}.} \label{fig2}
\end{figure}
\section*{Stripes Alter the phonon dispersion?}

That 1/8 doping anomalies are observed in both Y-123 and Hg cuprates suggests
they are a generic feature of underdoped CuO$_2$ planes.  In this context the
doping behavior of $\kappa$ for Hg-1223 [Fig. 2 (a)] is of interest since it
presumably approximates the phase behavior in the absence of localized holes.
We expect the electronic contribution, $\kappa_e$, to increase smoothly with
increasing $p$. The upturn in $\kappa$ above optimal doping ($p$=0.16) is
presumably attributable to this rising $\kappa_e$.  The most reliable estimate
of $\kappa_e$ in the cuprates comes from thermal Hall conductivity
measurements\cite{Krishana} on Y-123 which imply $\kappa_e/\kappa\simeq 0.1$
for in-plane heat conduction near optimal doping.  The data in Fig. 1 (a), (b)
suggest that this ratio is roughly the same in polycrystals, thus motivating
the dotted and dashed curves in Fig. 2 (a) as an educated guess for the
electronic and lattice terms, respectively, in Hg-1223. The lattice
conductivity, $\kappa_L=\kappa-\kappa_e$, predominates in the underdoped
regime and is peaked near $p$=1/8.

There is experimental support for the proposal that this peak in $\kappa$ is
associated with doping-dependent changes in the phonon dispersion which are
generic to underdoped cuprates.  Figure 2~(b) shows the behavior of the
normal-state, transverse shear elastic constant (proportional to the square of
the sound velocity) for single-crystal La-214\cite{LSCOelastic}, the only
material for which doping-dependent measurements for all the main symmetry
directions have been reported to our knowledge. A substantial hardening of the
lattice in the underdoped regime, with a maximum near $p=x\simeq$1/8, was
observed for all symmetries, indicating that the changes with doping are
systemic.

It is possible that the phase behavior of the elastic constants for La-214 and
$\kappa_L$ for Hg-1223 reflect a renormalization of the lattice dispersion due
to changes in stripe dynamics with doping. At $p$=1/8 the stripes are
maximally commensurate with the lattice, and their fluctuations should be
minimal. Enhanced fluctuations are to be expected at higher doping, due to the
destabilizing role of repulsive interactions in stripes that neutron
scattering results (Yamada {\it et al.}\cite{DynamicStripes}) suggest, are
charge compressed.  For $p<$1/8, the cluster spin-glass state observed by
$\mu$SR studies\cite{Niedermayer} is characterized by increasing magnetic
disorder with decreasing $p$ in the range $0.08 < p <0.12$, possibly
associated with increasing disorder\cite{Emery} in the stripe period.  It is
plausible that disorder in the stripe system at both higher and lower doping
about $p$=1/8 induces a softening of the lattice that is reflected in Fig.~2.
Theoretical investigations of elastic coupling to the stripe system would
certainly be of interest.

\begin{figure}[b!] 
\centerline{\epsfig{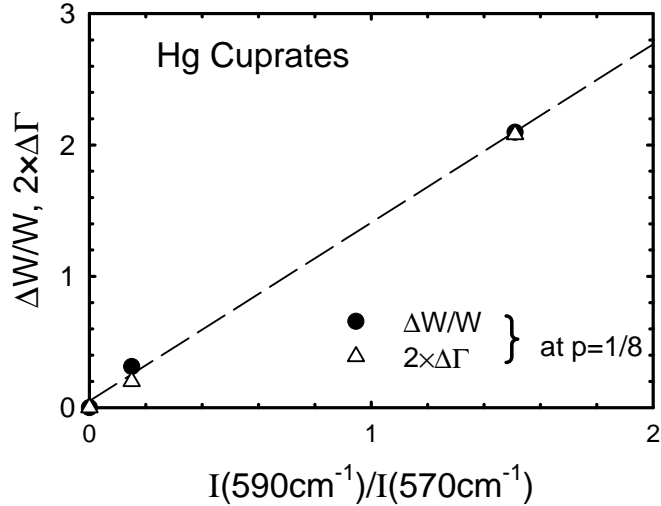}}
\vskip -3in \caption{(a) The enhancement in thermal resistivity
and suppression in $\Gamma$ for the Hg compounds at $p$=1/8,
relative to values for Hg-1223, plotted versus the oscillator
strength ratio of Raman-scattering defect modes
(Ref.~\protect\cite{HgRaman}).} \label{fig3}
\end{figure}
\section*{Oxygen Vacancies and 1/8 Anomalies}

Returning now to the enhancement of $W$ and suppression of $\Gamma$ near
$p$=1/8, our measurements suggest these phenomena reflect stripe pinning (or
reduced stripe fluctuations) associated with oxygen vacancy clusters.  Raman
studies of defect modes in the Hg materials\cite{HgRaman} provide a measure of
the density of vacancy clusters. The 590 cm$^{-1}$ Raman mode in Hg cuprates
is attributed to c-axis vibrations of apical oxygen in the presence of oxygen
vacancies on each of the four nearest-neighbor sites in the HgO$_{\delta}$
layers.  The amount by which the thermal resistivity and $\Gamma$ values of
Hg-1201 and Hg-1212 differ from those of Hg-1223 at $p$=1/8 both correlate
well with the integrated oscillator strength of this vibrational mode
normalized by that of the 570 cm$^{-1}$ mode (Fig.~\ref{fig3}), the latter
attributed to apical vibrations in the presence of fewer than four vacancies,
and common in the spectra\cite{HgRaman} of all three Hg materials.  We infer
that oxygen-vacancy clusters, of size four or more in the Hg cuprates, can pin
a stripe fragment.

Supporting the role of oxygen vacancies in the observed 1/8 features for Y-123
are our recent measurements on oxygen-stoichiometric
Y$_{1-x}$Ca$_x$-124\cite{CohnYCa124}. The $\Gamma$ values for $x$=0, 0.10,
0.15 are plotted in Fig.~1~(b).  There is no evidence for suppression of
$\Gamma$ near 1/8 doping, and the data follow the scaled $\mu$SR curve quite
well.  Within the context of the interpretation we have outlined, we conclude
that Ca substitution for Y (at the level of $\leq$ 15\%) is not as effective
in stripe pinning as are oxygen vacancy clusters.  It remains to be determined
by what mechanism these clusters induce pinning.

\section*{acknowledgements}
This work was supported by NSF Grant No. DMR-9631236.

\end{document}